\documentclass[amssymb,aps,prl,floatfix,twocolumn]{revtex4}
\usepackage[intlimits]{amsmath}
\usepackage{amssymb}
\usepackage{graphicx}
\usepackage{tensor}
\usepackage[dvipsnames]{xcolor}

\begin{document}

\author{Patryk Mach}\email{patryk.mach@uj.edu.pl}
\affiliation{Instytut Fizyki Teoretycznej, Uniwersytet Jagiello\'{n}ski, \L ojasiewicza 11,  30-348 Krak\'ow, Poland}
\author{Andrzej Odrzywo{\l}ek}\email{andrzej.odrzywolek@uj.edu.pl}
\affiliation{Instytut Fizyki Teoretycznej, Uniwersytet Jagiello\'{n}ski, \L ojasiewicza 11,  30-348 Krak\'ow, Poland}

\title{Accretion of Dark Matter onto a Moving Schwarzschild Black Hole: An Exact Solution}

\begin{abstract}
We investigate accretion of dark matter onto a moving Schwarzschild black hole. The dark matter is modeled by the collisionless Vlasov gas, assumed to be in thermal equilibrium at infinity. We derive an exact stationary solution and provide a compact formula for the mass accretion rate. In general, the mass accretion rate is a nonmonotonic function of the black hole velocity. A monotonic relation (the accretion rate proportional to the Lorentz factor associated with the velocity of the black hole) is obtained for high asymptotic temperatures of the gas. The derived accretion rates are relevant for the growth of primordial black holes in the early Universe.
\end{abstract}


\maketitle

\paragraph{Introduction}

The theory of accretion onto compact objects originated a century ago with the works of Hoyle, Lyttleton, and Bondi, investigating the in-fall of matter onto a compact object (a star) moving through a cloud of gas \cite{hoyle_lyttleton, lyttleton_hoyle, bondi_hoyle}. Newtonian models of this kind have been constructed for a variety of matter models, ranging from simple ballistic approximations to hydrodynamical solutions \cite{bisnovatyi} (see \cite{edgar} for a review). In contrast to that, relativistic models of the accretion onto a moving black hole are relatively new. The best known general-relativistic solution has been obtained for the so-called potential flows in \cite{petrich}; it corresponds to a stationary accretion of the perfect fluid with the ultrahard equation of state (the pressure being equal to the energy density) onto a moving Schwarzschild or Kerr black hole. A straightforward generalization of this result for the Kerr-Newman metric has been obtained in \cite{babichev}. A general-relativistic version of the ballistic approximation was derived in \cite{tajeda}. In the generic hydrodynamical case, solutions can be obtained numerically \cite{PSST, font, font_ibanez, font_ibanez_papadopoulos, donmez, zanotti, blakely, lora, cruz}. A review of these works can be found in \cite{rezzolla_zanotti}. 

In this Letter we present an exact solution representing accretion of the collisionless Vlasov gas onto a moving Schwarzschild black hole. A Newtonian model of the spherical accretion of the Vlasov gas has been derived by Zel'dovich and Novikov in \cite{zeldovich} and popularized by Shapiro and Teukolsky in \cite{shapiro_teukolsky}. The spherical accretion of the relativistic Vlasov gas onto a Schwarzschild black hole has recently been investigated by Rioseco and Sarbach, who developed a Hamiltonian formalism, allowing for an analysis of more complex flows on the fixed Schwarzschild background \cite{Olivier,Olivier2} (see also \cite{cieslik_mach}). Our analysis is based on this formalism. The essential features of the model discussed in this Letter are common with other approaches to the relativistic Bondi-Hoyle-Lyttleton accretion. We derive a stationary solution representing the motion of matter in a fixed Schwarzschild background, assuming the asymptotic conditions corresponding to the gas in a uniform motion along a fixed direction. More precisely, they correspond to the Maxwell-J\"{u}ttner distribution representing the gas in thermal equilibrium, boosted along a fixed direction with a constant speed. A detailed analysis and technical aspects of our derivation can be found in the accompanying paper \cite{duzy}.

Nowadays, the relativistic Vlasov gas can be understood as a model for a noninteracting dark matter---see, for example, \cite{DM_Vlasov} for a Vlasov model of dark matter halos. On the other hand, the accretion of the dark matter on black holes is usually modeled either with nonrelativistic approximations \cite{read} or by referring to known perfect-fluid solutions \cite{peirani}. Our result fills this gap by providing an exact relativistic solution describing the accretion of the collisionless gas. Although the description of the morphology of the flow is quite involved, we obtain a remarkably compact formula for the mass accretion rate. It should be relevant for the accretion of dark matter onto black holes passing through galactic halos and for the modeling of the growth of primordial black holes in the early Universe (cf.\ \cite{zeldowicz, bicknell1, bicknell2, ricotti,ostriker}).

\paragraph{Vlasov gas in the Schwarzschild spacetime}

In terms of the horizon-penetrating Eddington-Finkelstein coordinates, the Schwarzschild metric can be written as
\begin{eqnarray}
g & = & - N dt^2 + 2(1 - N) dt dr + (2 - N) dr^2 \nonumber \\
&& + r^2 (d \theta^2 + \sin^2 \theta d \varphi^2),
\label{efgeneral}
\end{eqnarray}
where $N = 1 - 2M/r$. We use geometric units with $c = G = 1$, where $c$ is the speed of light, and $G$ denotes the gravitational constant.

We will work in the Hamiltonian formulation. The Hamiltonian of a particle traveling along a timelike geodesic $\Gamma$ can be written as
\begin{eqnarray}
H(x^\mu,p_\nu) & = & \frac{1}{2}g^{\mu \nu}(x^\alpha) p_\mu p_\nu \nonumber \\
& = & \frac{1}{2} \left[ g^{tt}(r) p_t^2 + 2 g^{tr}(r) p_t p_r + g^{rr}(r) p_r^2 \right. \nonumber \\
&& \left. + \frac{1}{r^2} \left( p_\theta^2 + \frac{p_\varphi^2}{\sin^2 \theta} \right) \right],
\label{hamiltonian_schw}
\end{eqnarray}
where $(x^\mu,p_\mu)$ are understood as canonical variables. It is easy to show that Hamilton's equations imply the well-known geodesic equation. In the following, we require that $H = 
- \frac{1}{2}m^2$, where $m$ denotes the rest mass of the particle. By standard arguments, $E = -p_t$, $l_z = p_\varphi$, $m = \sqrt{-2H}$, and $l = \sqrt{p_\theta^2 + p_\varphi^2/\sin^2 \theta}$ are constants of motion. 

The Vlasov gas is described in terms of the phase-space distribution function $f = f(x^\mu,p_\nu)$. The condition that $f$ should remain constant along a geodesic leads to the Vlasov equation, which we write as
\begin{equation}
\label{vlasoveq}
\frac{\partial H}{\partial p_\mu} \frac{\partial f}{\partial x^\mu} - \frac{\partial H}{\partial x^\nu} \frac{\partial f}{\partial p_\nu} = 0.
\end{equation}
The particle current density and the energy-momentum tensor can be computed as
\begin{equation}
\label{jgeneral}
J_\mu = \int p_\mu f (x,p) \mathrm{dvol}_x(p)
\end{equation}
and
\begin{equation*}
T_{\mu\nu} = \int p_\mu p_\nu f (x,p) \mathrm{dvol}_x(p),
\end{equation*}
where
\begin{equation}
\mathrm{dvol}_x(p) = \sqrt{- \mathrm{det}[g^{\mu\nu}(x)]} dp_0dp_1dp_2dp_3.
\end{equation}
Equation (\ref{vlasoveq}) can be used to show that $\nabla_\mu J^\mu = 0$ and $\nabla_\mu T^{\mu \nu} = 0$, where $\nabla_\mu$ denotes the covariant derivative associated with the metric $g$. The particle number density can be defined covariantly as $n = \sqrt{-J_\mu J^\mu}$.

The solution of the Vlasov equation is based on the following construction of new phase-space coordinates $(Q^\mu,P_\nu)$. We start by considering the motion along a geodesic $\Gamma$ parametrized by constants $E$, $l_z$, $l$, and $m$ and define the abbreviated action
\begin{equation}
S = \int_\Gamma p_\mu dx^\mu = -E t + l_z \varphi + \int_\Gamma p_r dr + \int_\Gamma p_\theta d\theta,
\end{equation}
which is used as a generating function for the canonical transformation $(t,r,\theta,\varphi,p_t,p_r,p_\theta,p_\varphi) \to (Q^\mu,P_\nu)$. New momenta $P_\mu$ are defined as $P_0 = m$, $P_1 = E$, $P_2 = l_z$, and $P_3 = l$, and the corresponding conjugate variables are $Q^0 = \partial S/\partial m$, $Q^1 = \partial S/\partial E$, $Q^2 = \partial S/\partial l_z$, and $Q^3 = \partial S/\partial l$. In practice, we will only need an explicit expression for $Q^3$, which can be written as
\begin{equation}
Q^3 =- l \int_\Gamma \frac{dr}{r^2 \left( -g^{tr} E + g^{rr} p_r \right)} + l \int_\Gamma \frac{d \theta}{p_\theta}.
\end{equation}

New coordinates $(Q^\mu,P_\nu)$ can be used to solve the Vlasov equation. The left-hand side of the Vlasov equation (\ref{vlasoveq}), which can be written as the Poisson bracket $\{ H, f \}$, is covariant with respect to canonical transformations. Since $H = -P_0^2/2$, Eq.\ (\ref{vlasoveq}) can be written as $\partial f /\partial Q^0 = 0$. Consequently, any distribution function $f$ independent of $Q^0$ satisfies the Vlasov equation.

In this Letter we consider a stationary and axially symmetric solution on a fixed Schwarzschild background. It was shown in \cite{Olivier} that such a solution should also satisfy $\partial f/\partial Q^1 = 0$ and $\partial f/\partial Q^2 = 0$.  In the following, we will search for a global solution of the form
\begin{equation}
f(x^\mu, p_\nu) = \mathcal F (Q^3, P_0, P_1, P_2, P_3).
\end{equation}

In the remainder of this Letter we will mostly use dimensionless quantities introduced in \cite{Olivier}. We define $r = M \xi$, $p_r = m \pi_\xi$, $p_\theta = M m \pi_\theta$, $E = m \varepsilon$, $l = M m \lambda$, and $l_z = M m \lambda_z$. The dimensionless momenta $\pi_\xi$ and $\pi_\theta$ can be expressed as
\begin{equation}
\label{pixi}
\pi_\xi = \frac{(1 - N \eta) \varepsilon + \epsilon_r \sqrt{\varepsilon^2 - U_\lambda(\xi)}}{N},
 \end{equation}
and
\begin{equation}
\label{pitheta}
\pi_\theta = \epsilon_\theta \sqrt{\lambda^2 - \frac{\lambda_z^2}{\sin^2 \theta}},
\end{equation}
where we have introduced the signs $\epsilon_r = \pm 1$, $\epsilon_\theta = \pm 1$, corresponding to the directions of motion along $\Gamma$, and the dimensionless effective potential reads
\begin{equation}
U_\lambda(\xi) = N \left(1 + \frac{\lambda^2}{\xi^2} \right) = \left(1 - \frac{2}{\xi} \right) \left(1 + \frac{\lambda^2}{\xi^2} \right).
\end{equation}

For $Q^3$ we obtain
\begin{equation}
Q^3 = \epsilon_r X(\xi,\varepsilon,\lambda) - \epsilon_r \frac{\pi}{2} - \epsilon_\theta \arctan \left( \frac{\lambda \cot \theta}{\sqrt{\lambda^2 - \frac{\lambda_z^2}{\sin^2 \theta}}} \right),
\end{equation}
where
\begin{equation}
X(\xi,\varepsilon,\lambda) = \lambda \int_\xi^\infty \frac{d \xi^\prime}{{\xi^\prime}^2 \sqrt{\varepsilon^2 - U_\lambda(\xi^\prime)}}.
\label{X}
\end{equation}
The above integral can be expressed in terms of elliptic functions.

\paragraph{Boosted Maxwell-J\"{u}ttner distribution}

We will construct our solutions assuming that the gas is asymptotically in thermal equilibrium and moves uniformly with a given velocity $v$. In practice, we assume that it is described by a boosted Maxwell-J\"{u}ttner distribution.

Consider the flat Minkowski spacetime with the metric
\begin{equation}
g = -dt^2 + dr^2 + r^2(d\theta^2 + \sin^2 \theta d \varphi^2).
\end{equation}
The Maxwell-J\"{u}ttner distribution corresponding to a simple, i.e., composed of same-mass particles, nondegenerate gas, boosted with a constant velocity along the axis ($\theta = 0, \theta = \pi$) can be written as
\begin{equation}
\label{juttnera}
f(x^\mu,p_\nu) = \delta\left( \sqrt{-p_\mu p^\mu} - m \right) F(x^\mu,p_\nu),
\end{equation}
where
\begin{equation}
\label{boostedf}
F (x^\mu,p_\nu) = \alpha \exp \left\{ \frac{\beta}{m} \gamma \left[ p_t - v \left( \cos \theta p_r - \frac{\sin \theta}{r} p_\theta \right) \right] \right\}
\end{equation}
and the Dirac delta enforces the mass shell condition. Here $\gamma = (1 - v^2)^{-1/2}$ is the Lorentz factor associated with the velocity $v$. The constant $\beta$ is related to the temperature $T$ of the gas at rest by $\beta = m/(k_\mathrm{B} T)$, where $k_\mathrm{B}$ denotes the Boltzmann constant.

The particle number density corresponding to this distribution is constant and independent of $v$. It reads $n_\infty = 4 \pi \alpha m^4 K_2(\beta)/\beta$, where $K_2$ is the modified Bessel function of the second kind \cite{israel}.

It turns out that the analysis of the asymptotic expressions is sufficient for our purposes.  For $r \to \infty$, the function $F$ given by Eq.\ (\ref{boostedf}) tends to
\begin{equation}
F = \alpha \exp \left[ \frac{\beta}{m} \gamma \left( p_t - v \cos \theta p_r \right) \right].
\label{boostedfasymptot}
\end{equation}

Constructing the coordinates $(Q^\mu,P_\nu)$ in the flat Minkowski spacetime is simple. The momenta $P_\mu$ are given by the same formulas as in in the Schwarzschild spacetime. Out of conjugate coordinates $Q^\mu$, we only need $Q^3$, which is given by
\begin{equation}
Q^3 = - \epsilon_r \frac{\pi}{2} - \epsilon_\theta \arctan \left( \frac{l \cot \theta}{\sqrt{l^2 - \frac{l_z^2}{\sin^2 \theta}}} \right),
\end{equation}
and where $\epsilon_r$ and $\epsilon_\theta$ have the analogous meaning to the Schwarzschild case, as defined in Eqs.\ (\ref{pixi}) and (\ref{pitheta}). 

A straightforward calculation allows one to express the asymptotic distribution (\ref{boostedfasymptot}) as
\begin{widetext}
\begin{equation}   
\label{factionangle}
F = \alpha \exp \left\{ \frac{\beta}{P_0} \gamma \left[ -P_1 - v \epsilon_r \epsilon_\theta \sqrt{P_1^2 - P_0^2} \frac{\sqrt{P_3^2 - P_2^2}}{P_3} \sin \left( -Q^3 - \epsilon_r \frac{\pi}{2} \right) \right]\right\}.
\end{equation}
\end{widetext}
By construction, expression (\ref{factionangle}) represents the global solution, valid also in the Schwarzschild spacetime.

\paragraph{Particle current density}

\begin{figure}
\includegraphics[width=\columnwidth]{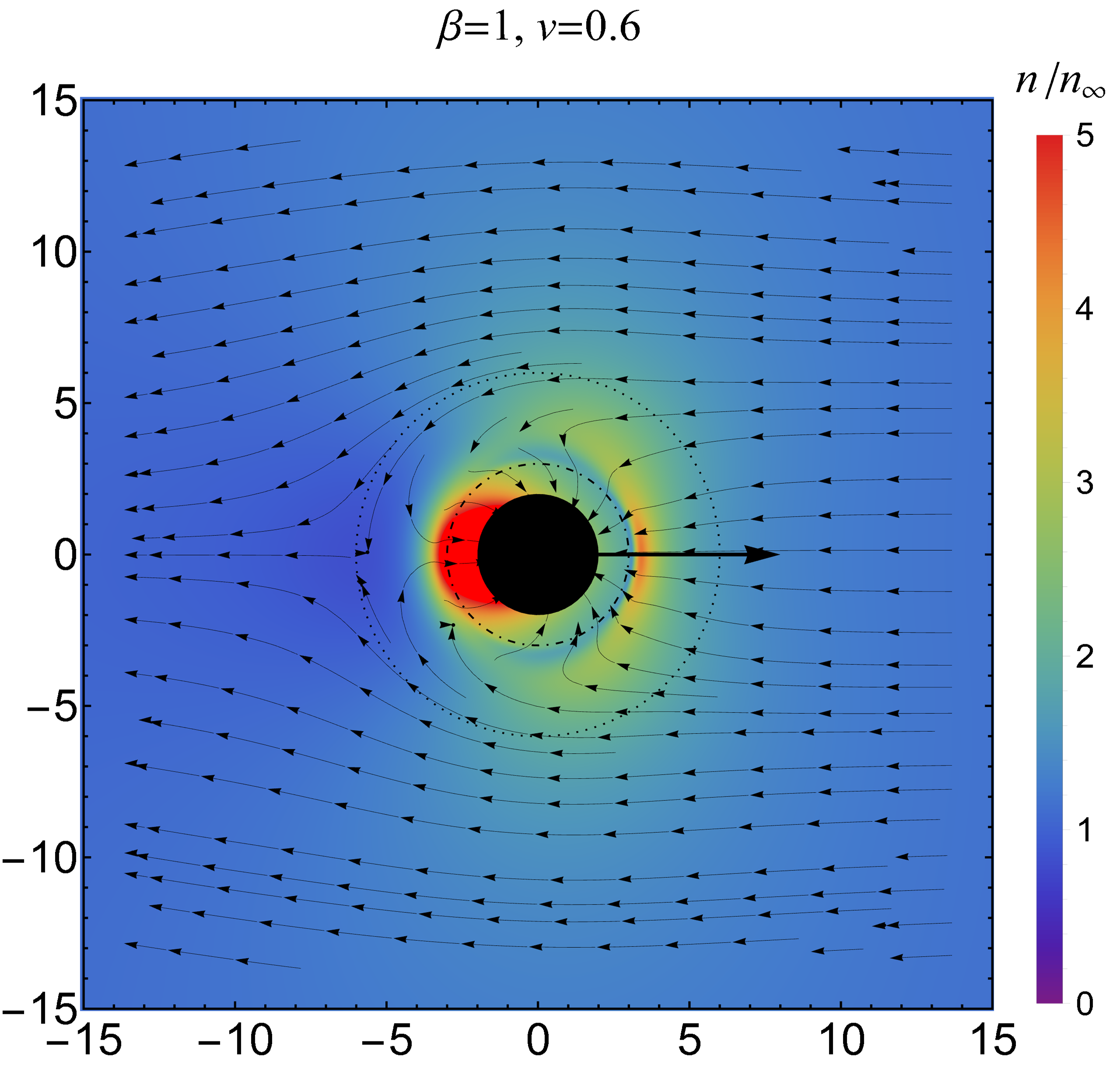}
\caption{\label{figure} 
Sample morphology of the flow in the vicinity of the black hole for $\beta = 1$, $v= 0.6$. The color scale depicts the the ratio $n/n_\infty$. The streamlines show the directions of the spatial part of the particle density current.}
\end{figure}

To compute the momentum integrals in (\ref{jgeneral}) one needs to control the phase-space region of integration. This can be done by analyzing the properties of the effective potential $U_\lambda(\xi)$ and by introducing yet another set of coordinates. We restrict the discussion to the particle trajectories that originate at infinity and get absorbed by the black hole [the quantities referring to these trajectories are denoted with (abs)] and the trajectories which also originate at infinity, but whose angular momentum is sufficiently high, so that they are scattered by the black hole
[the quantities referring to these particles are denoted with (scat)]. The phase-space regions occupied by these two classes of particles are characterized in \cite{Olivier}. Here we only summarize the results. 

One starts with defining the critical angular momentum
\begin{equation}
\label{limit1}
 \lambda_c(\varepsilon)^2 = \frac{12}{1-\frac{4}{\left(\frac{3 \varepsilon }{\sqrt{9 \varepsilon ^2-8}}+1\right)^2}}.
 \end{equation}
The region in the phase space occupied by the absorbed particles is characterized by  $\varepsilon \ge 1$ and $\lambda \le \lambda_c(\varepsilon)$.

A minimal energy of a scattered particle is given by
\begin{equation}
\label{limit2}
    \varepsilon_\mathrm{min}(\xi) = \begin{cases} \infty, & \xi \leq 3 ,\\
    \sqrt{\left(1 - \frac{2}{\xi} \right) \left(1 + \frac{1}{\xi - 3} \right)}, & 3 < \xi < 4, \\
    1, & \xi \ge 4.  \end{cases}
\end{equation}
Note that no scattered particles can be found below the photon sphere ($\xi \le 3$). At the same time, the upper limit on the total angular momentum reads
\begin{equation} 
\label{limit3}
\lambda_\mathrm{max}(\xi,\varepsilon) = \xi \sqrt{\frac{\varepsilon^2}{1 - \frac{2}{\xi}} - 1}. 
\end{equation}
The phase-space region occupied by the particles traveling from infinity and scattered off the centrifugal barrier is given by the conditions $\varepsilon_\mathrm{min}(\xi) < \varepsilon < \infty$, $\lambda_c(\varepsilon) < \lambda < \lambda_\mathrm{max}(\xi,\varepsilon)$.

The integrals over momenta in Eq.\ (\ref{jgeneral}) can be computed conveniently with the help of new momentum coordinates $(\varepsilon,m,\lambda,\chi)$, where $\chi$ is defined by
\begin{equation}
\pi_\theta = \lambda \cos \chi, \quad \lambda_z = \lambda \sin \theta \sin \chi.
\end{equation}
For the integration element in the momentum space we get
\begin{equation}
\mathrm{dvol}_x(p) = \frac{1}{\xi^2} \frac{m^3 \lambda}{ \sqrt{\varepsilon^2 - U_\lambda (\xi)}} d \varepsilon dm d\lambda d \chi.
\end{equation}
In terms of these new coordinates, $F$ can be written as
\begin{equation}
F = \alpha \exp(-\beta \gamma \varepsilon) \exp(- \epsilon_r Y) \exp(Z \cos \chi),
\end{equation}
where
\begin{subequations}
\label{YZ}
\begin{eqnarray}
\label{Y}
Y & = & \beta \gamma v \sqrt{\varepsilon^2 - 1} \cos X (\xi, \varepsilon, \lambda) \cos \theta, \\
\label{Z}
Z & = & \beta \gamma v \sqrt{\varepsilon^2 - 1} \sin X (\xi, \varepsilon, \lambda) \sin \theta.
\end{eqnarray}
\end{subequations}

The components of $J_\mu$ are computed by separating them into two parts, $J_\mu = J_\mu^\mathrm{(abs)} + J_\mu^\mathrm{(scat)}$, which correspond to absorbed and scattered particles, respectively. The components $J_\mu^\mathrm{(abs)}$ can be expressed as
\begin{equation}
J_\mu^\mathrm{(abs)} = \int_1^\infty d\varepsilon \int_0^{\lambda_c(\varepsilon)} d \lambda \int_0^{2 \pi} d \chi \int_0^\infty dm \frac{ p_\mu f m^3 \lambda}{ \xi^2 \sqrt{\varepsilon^2 - U_\lambda (\xi)}},
\end{equation}
where we take $\epsilon_r = -1$ in the expressions for $p_r$ and $f$. The components $J_\mu^\mathrm{(scat)}$ are computed as
\begin{eqnarray}  J_\mu^\mathrm{(scat)} & = & \sum_{\epsilon_r = \pm 1} \int_{\varepsilon_\mathrm{min}(\xi)}^\infty d\varepsilon \int_{\lambda_c(\varepsilon)}^{\lambda_\mathrm{max}(\xi,\varepsilon)} d \lambda \int_0^{2 \pi} d \chi \int_0^\infty dm  \nonumber \\
&& \times \frac{p_\mu f m^3 \lambda}{ \xi^2 \sqrt{\varepsilon^2 - U_\lambda (\xi)}}.
\end{eqnarray}

In explicit terms, the formulas for $(J_\mu) = (J_t, J_r, J_\theta, J_\varphi)$ are as follows:
\begin{widetext}
\begin{subequations}
\label{jrjtheta}
\begin{eqnarray}
\left( J_\mu^\mathrm{(abs)} \right) & = &  \frac{2 \pi \alpha m^4}{\xi^2} \int_1^\infty d \varepsilon e^{- \beta \gamma \varepsilon}
\int_0^{\lambda_c(\varepsilon)} d \lambda \frac{\lambda \exp(Y)}{\sqrt{\varepsilon^2 - U_\lambda (\xi)}} \\
&& \times 
\left\{   - \varepsilon I_0(Z) , I_0(Z)  \left( - \varepsilon + \frac{1 + \frac{\lambda^2}{\xi^2}}{\varepsilon + \sqrt{\varepsilon^2 - U_\lambda (\xi)}} \right) , M \lambda I_1 (Z) , 0 \right\}, \\
\left( J_\mu^\mathrm{(scat)} \right) & = & \frac{4 \pi \alpha m^4}{\xi^2} \int_{\varepsilon_\mathrm{min}(\xi)}^\infty d \varepsilon e^{-\beta \gamma \varepsilon} \int_{\lambda_c(\varepsilon)}^{\lambda_\mathrm{max}(\xi,\varepsilon)} d \lambda \frac{\lambda \cosh{(Y)}}{\sqrt{\varepsilon^2 - U_\lambda (\xi)}} \\
& &
\times \left\{   
- \varepsilon I_0(Z), 
\frac{I_0(Z)}{N}  \left[ 
(1 - N) \varepsilon - \sqrt{\varepsilon^2 - U_\lambda (\xi)} \tanh (Y) \right], - M \lambda I_1 (Z) , 0 \right\},
\end{eqnarray}
\end{subequations}
\end{widetext}
where $Y$ and $Z$ are given by Eqs.\ (\ref{YZ}), and $I_n$ denotes the modified Bessel function of the first kind. Note that $J_\varphi = 0$.

Numerical evaluation of integrals in Eqs.\ (\ref{jrjtheta}) is not exactly straightforward, and it is time consuming, mostly because the formula for $X(\xi,\varepsilon,\lambda)$ is given in terms of elliptic functions. An example of the morphology of the flow near the black hole moving with the speed $v = 0.6$ is shown in Fig.\ \ref{figure}.

\paragraph{Mass accretion rate}

Another key observable is the mass accretion rate through a sphere of a given radius $r = M \xi$. It can be defined as
\begin{equation} 
\label{Mdot_general}
\dot M = - m \int_0^{2 \pi} d\varphi \int_0^\pi d \theta\; r^2 \sin \theta \; J^r. 
\end{equation}

As a consequence of the continuity equation, $\dot M$ does not depend on the radius of the sphere. Choosing $r \to \infty$, we get
\begin{eqnarray}
\label{Mdot_infty}
\dot M & = & \pi G^2 c^{-3} M^2 m n_\infty \frac{\beta}{K_2(\beta)}\\
&& \times \int_1^\infty d \varepsilon \; e^{-\beta \gamma \varepsilon} \lambda_c^2(\varepsilon) \frac{\sinh \left( \beta \gamma v c^{-1} \sqrt{\varepsilon^2 - 1}  \right)}{\beta \gamma v c^{-1} \sqrt{\varepsilon^2 - 1}}, \nonumber
\end{eqnarray}
where we have returned to cgs-SI units. Note that the term $\pi G^2 c^{-3} M^2 m n_\infty$ is directly related to the Hoyle-Lyttleton mass accretion rate $\dot M_\mathrm{HL} = 4 \pi G^2 M^2 m n_\infty/v^3$ \cite{hoyle_lyttleton} evaluated for $v = c$.

For $v^2 \ll 3/\beta$, we get the following low-temperature ($\beta \to \infty$) limit:
\begin{equation}
\dot{M} \simeq  16 G^2 c^{-3} M^2 m n_\infty \sqrt{2 \pi \beta}
\left( 1 - \beta v^2/6 \right).
\end{equation}
For $v=0$ this result has already been obtained in \cite{Olivier}. For $\beta \to 0$, one obtains
\begin{equation}
\label{Mdot_limit}
\dot M = 27 \pi G^2 c^{-3} M^2 m  n_\infty  \gamma; 
\end{equation}
i.e., $\dot M$ is directly proportional to $\gamma$. The same qualitative behavior was also observed in \cite{petrich} for the ultrahard perfect fluid. In general, $\dot M$ does not have to be a monotonic function of $v$. For $\beta \gtrapprox 5$, the accretion rate decreases with the black hole speed for nonrelativistic values of $v$; it attains a minimum located around $v \approx 0.6 c$ and grows for ultrarelativistic values of $v$, roughly proportionally to $\gamma$. For $\beta = 100$, the accretion rate $\dot M$ corresponding to $v = 0.5 c$ is roughly one-third of the value corresponding to $v = 0$. The value given by Eq.\ (\ref{Mdot_limit}) is a lower limit of $\dot M$, computed with Eq.\ (\ref{Mdot_infty}). An upper limit of $\dot M$ and an approximation for $\beta \to \infty$ can be obtained as
\begin{equation}
\label{binfty}
\dot M = \pi G^2 c^{-3} M^2 m  n_\infty \frac{\lambda_c(\gamma)^2}{\gamma v}. 
\end{equation}

The energy accretion rate $\dot {\mathcal E}$ can be defined in way analogous to $\dot M$, basically by replacing the current $m J^\mu$ in Eq.\ (\ref{Mdot_general}) with $-T\indices{^\mu_t}$. For $\beta \to 0$, one obtains
\begin{equation}
\label{Edot_limit}
\dot {\mathcal E} = 36 \pi G^2 c^{-3} M^2 \varepsilon_\infty \left( \gamma^2 - \frac{1}{4} \right),
\end{equation}
where $\varepsilon_\infty$ is the asymptotic energy density \cite{duzy}.

While a detailed analysis of implications of Eqs.\ (\ref{jrjtheta}) and (\ref{Mdot_infty}) is beyond the scope of this Letter, we would like to give some remarks related to a possible growth of black holes in the early Universe, much in the spirit of Zel'dovich and Novikov \cite{zeldowicz}. One of characteristic properties of Eqs.\ (\ref{Mdot_infty})--(\ref{binfty}) is an existence of a finite time $t_1$ at which $M \to \infty$, assuming a literal understanding of these formulas as $dM/dt \propto M^2$. Of course, Eqs.\ (\ref{Mdot_infty})--(\ref{binfty}) are only valid for a stationary process, we do not take into account that the black hole accumulating matter slows down, etc. Consequently, the blowup time $t_1$ can only be understood as a characteristic timescale associated with the accretion process. Moreover, a blowup-type behavior seems to be unphysical for $v = 0$, because of the exhaustion of the reservoir of particles. This might not be the case for moving black holes, which could sweep through new matter reservoirs and grow, in principle, indefinitely. Unfortunately, the time $t_1$ in the present Universe is by many orders of magnitude larger than its age, even for supermassive black holes and realistic matter density contrasts. The situation seems to be different for primordial black holes created in early epochs. In the standard Friedmann-Lema\^{\i}tre-Robertson-Walker universe filled with hot dark matter particles, the energy density would decrease as $\varepsilon_\infty = 3 c^2/(32 \pi G t^2)$ [an expression valid for the radiation-dominated universe and $t \to 0$, cf.\ \cite{LandauLifszyc}, Eq.\ (112.15)]. Assuming the black hole growth rate $dM/dt \approx \dot{\mathcal E}/c^2$, we get from Eq.\ (\ref{Edot_limit})
\begin{equation}
\frac{dM}{dt} = \frac{27}{8} G c^{-3} \left( \gamma^2 - \frac{1}{4} \right) M^2 t^{-2}.
\end{equation}
Integrating from time $t_0$ to $t$ we obtain
\begin{equation}
M = \left[ \frac{1}{M_0} + \frac{27G}{8c^3}\left(\gamma^2 - \frac{1}{4} \right)\left(\frac{1}{t} - \frac{1}{t_0}\right) \right]^{-1}
\end{equation}
for the accretion process that starts at time $t_0$ with a black hole of mass $M_0$. Provided that $M_0 (\gamma^2-1/4) > 8 c^3 t_0 /(27 G )$, the blowup occurs for
\begin{equation}
t_1 = \left[ \frac{1}{t_0} - \frac{8 c^3}{27 G M_0 (\gamma^2-1/4)} \right]^{-1}.
\end{equation}
Assuming further $M_0 = \kappa c^3 t_0/G$ \cite{zeldowicz,carr}, where $\kappa$ is a free parameter, we get finally
\begin{equation}
\frac{t_0}{t_1} = 1 - \frac{8}{27 \kappa (\gamma^2 - 1/4)}.
\end{equation}
For $\kappa \approx 1$ and $\gamma \approx 1$, this yields $t_0/t_1 \approx 0.6$, suggesting that the accretion could be a robust process in the early Universe.

\paragraph{Acknowledgments} P.\ M.\ was partially supported by the Polish  National Science Centre Grant No.\ 2017/26/A/ST2/00530.

\end{document}